\documentclass[12pt]{article}
\pdfoutput=1
\usepackage{myjheppub}
\usepackage{graphicx,booktabs,latexsym,amssymb,amsmath,bm,epsfig,psfrag,float}
\hypersetup{pdfnewwindow=true}

\usepackage{bbm}
\usepackage{mathtools}
\usepackage{mathrsfs}
\usepackage{hepunits}
\usepackage{color}
\usepackage{multirow}
\usepackage{sectsty}
\allsectionsfont{\boldmath}
\allowdisplaybreaks

\usepackage[utf8]{inputenc}

\newcommand{\Tr}{\mathrm{Tr}}

\preprint{
\begin{flushright}
Nikhef 2018-057
\\
IPPP/18/99
\end{flushright}
}

\title{ Resummation for rapidity distributions in top-quark pair production }	

\author[a]{Benjamin~D.~Pecjak,}
\emailAdd{ben.pecjak@durham.ac.uk}

\author[b,c]{Darren J.~Scott,}
\emailAdd{d.j.scott@uva.nl}

\author[d]{Xing Wang,}
\emailAdd{x.wong@pku.edu.cn}

\author[d,e,f]{and Li Lin Yang}
\emailAdd{yanglilin@pku.edu.cn}

\affiliation[a]{Institute for Particle Physics Phenomenology, University of Durham, DH1 3LE Durham, UK}

\affiliation[b]{Institute for Theoretical Physics, University of Amsterdam, Science Park 904, 1098 XH Amsterdam, The Netherlands}
\affiliation[c]{Nikhef, Theory Group, Science Park 105, 1098 XG, Amsterdam, The Netherlands}

\affiliation[d]{School of Physics and State Key Laboratory of Nuclear Physics and Technology, Peking University, Beijing 100871, China}

\affiliation[e]{Collaborative Innovation Center of Quantum Matter, Beijing, China}

\affiliation[f]{Center for High Energy Physics, Peking University, Beijing 100871, China}

\abstract{We extend our framework for the simultaneous resummation of soft and small-mass logarithms to  rapidity distributions in top quark pair production. We give numerical results for the rapidity distribution of the top quark or the anti-top quark, as well as the rapidity distribution of the $t\bar{t}$ pair, finding that resummation effects stabilize the dependence of the differential cross sections on the choice of factorization scale. We compare our results with recent measurements at the Large Hadron Collider and find good agreement. Our results may be useful in the extraction of the gluon parton distribution function from $t\bar{t}$ production.}

\begin{document}

\maketitle

\section{Introduction}
\label{sec:intro}

Being the heaviest fundamental particle in the Standard Model (SM), the top quark plays an important role in studying the spontaneous breakdown of electroweak symmetry and in searching for new physics beyond the SM. The Large Hadron Collider (LHC) is expected to produce billions of top quarks during its lifetime and with such a large amount of data, precision studies in the top quark sector have become a key goal of the LHC program. Accurate measurements of the production and decay channels of the top quark allow us to probe some of the less-well-determined SM parameters, such as the CKM matrix element $V_{tb}$ as well
as the gauge and Yukawa coupling of the top quark, which are all sensitive to new physics effects. In addition, the kinematic distributions in top quark production may be affected by the existence of new resonances, particularly in the high energy tails. Last but not least, top quark production processes are important backgrounds for many processes in and beyond the SM, which need to be carefully modeled in order to extract possible new physics signals. 

Even if no deviations from the SM are found in the top quark sector, the differential cross sections in top quark pair production can still help to constrain the parton distribution function (PDF) of the gluon \cite{Czakon:2013tha, Gauld:2013aja, Czakon:2016olj}. This is due to the fact that around 90\% of the $t\bar{t}$ events at the LHC come from the gluon-initiated partonic subprocess. The precise knowledge of the gluon PDF is indispensable for Higgs physics, since at the LHC the Higgs boson is also mainly produced via the gluon-fusion process. The theoretical predictions for single Higgs boson production, Higgs boson production associated with a jet and Higgs boson pair production can be improved by incorporating the $t\bar{t}$ data in the PDF fit.

To match the high precision of the LHC experiments, it is necessary to have theoretical predictions with equally high or even higher accuracy. For differential cross sections in top quark pair production, the best fixed-order results are the next-to-next-to-leading order (NNLO) ones given in \cite{Czakon:2016dgf}. Combining with the resummation framework developed in \cite{Ahrens:2010zv, Ferroglia:2012ku} at next-to-next-to-leading logarithmic accuracy (NNLL$'$), the works \cite{Pecjak:2016nee, Czakon:2018nun} presented predictions for the $t\bar{t}$ invariant mass distribution and the top quark transverse momentum distribution up to the NNLO+NNLL$'$ precision. In this short article, we extend the resummation framework to the rapidity distribution of the top/anti-top quark, as well as the rapidity distribution of the $t\bar{t}$ pair. We present NNLO+NNLL$'$ predictions for these observables at the LHC with a center-of-mass energy $\sqrt{s}=\unit{13}{\TeV}$.

The paper is organized as follows. In section~\ref{sec:formalism} we briefly review the formalism for the resummation of threshold logarithms and for the joint resummation of threshold and small-mass logarithms. We then discuss the modifications necessary to our resummation formalism for the computation of rapidity distributions. In section~\ref{sec:results}, we present numerical predictions, with emphasis on the sensitivity of the results to  scale choices and to PDFs. We conclude in section~\ref{sec:conclusions}. In appendix~\ref{sec:Revised} we discuss a few differences between the formulas used in this work and in \cite{Pecjak:2016nee, Czakon:2018nun} due to a recent calculation of the NNLO soft function for top quark pair production~\cite{Wang:2018vgu}, and show the numerical impact of these corrections.

\section{Formalism}
\label{sec:formalism}
	
In this section, we briefly review the resummation framework established in \cite{Ahrens:2010zv, Ferroglia:2012ku, Czakon:2018nun}, and discuss its extension to describe rapidity distributions in $t\bar{t}$ production. We consider inclusive top quark pair production at the LHC
\begin{align}
p(P_1) + p(P_2) \to t(p_3) + \bar{t}(p_4) + X \, ,
\end{align}
where the differential cross section with respect to the $t\bar{t}$ invariant mass $M_{t\bar{t}}=(p_3+p_4)^2$ and the scattering angle $\theta$ in the center-of-mass frame can be written as
\begin{equation}
\frac{d^2\sigma(\tau)}{dM_{t\bar{t}}\,d\cos\theta} = \frac{8\pi\beta_t}{3sM_{t\bar{t}}} \sum_{ij} \int dz dx_1 dx_2 \, \delta(\tau-zx_1x_2) \, f_{i/p}(x_1,\mu_f) \, f_{j/p}(x_2,\mu_f) \, C_{ij}(z,\mu_f) \, .
\label{eq:x-sec1}
\end{equation}
Here we have defined $s=(P_1+P_2)^2$, $\tau=M_{t\bar{t}}^2/s$, and $\beta_t = \sqrt{1-4m_t^2/M_{t\bar{t}}^2}$. The hard-scattering kernel $C_{ij}(z,\mu_f)$ and the PDFs $f_{i/p}(x,\mu_f)$ depend on the factorization scale $\mu_f$. The sum runs over all parton species $i,j=q,\bar{q},g$. Note that we have suppressed the dependence of $C_{ij}$ on the kinematic variables $M_{t\bar{t}}$, $m_t$ and $\cos\theta$ for convenience. As in \cite{Czakon:2018nun}, we transform to Mellin space, where the convolution in eq.~(\ref{eq:x-sec1}) becomes a product which we write as
\begin{align}
\frac{d^2\tilde{\sigma}(N)}{dM_{t\bar{t}}\,d\cos\theta} = \frac{8\pi\beta_t}{3sM_{t\bar{t}}} \sum_{ij} \tilde{f}_{i/p}(N,\mu_f) \, \tilde{f}_{j/p}(N,\mu_f) \, \tilde{c}_{ij}(N,\mu_f) \, ,
\label{eq:x-sec2}
\end{align}
where the functions with tildes are the Mellin transforms of the corresponding momentum-space functions in eq.~(\ref{eq:x-sec1}), and $N$ is the Mellin moment variable.

In general, the hard-scattering kernels $C_{ij}(z,\mu_f)$ or $\tilde{c}_{ij}(N,\mu_f)$ can be calculated in perturbation theory as a series in the strong coupling $\alpha_s$. However, in certain kinematic configurations, the differential cross sections are dominated by soft emissions. In such cases, the hard-scattering kernels in the flavor diagonal channels (i.e., $ij=q\bar{q},gg$) are enhanced by Sudakov double logarithms $\ln^2N$ at each order in the strong coupling $\alpha_s$. The methods to resum these logarithms to all orders in $\alpha_s$ for $t\bar{t}$ production are well-known \cite{Kidonakis:1997gm, Ahrens:2010zv}. The resummation is based on the factorization of the cross section in the limit where final state gluons are soft and leads to the formula
\begin{multline}
\label{eq:soft-fact}
\widetilde{c}_{ij}(N,M_{t\bar{t}},m_t,\cos\theta,\mu_f) =
\Tr \Bigg[ \bm{H}^m_{ij}(M_{t\bar{t}},m_t,\cos\theta,\mu_f) \\*
\times \widetilde{\bm{s}}^m_{ij}\left(\ln\frac{M_{t\bar{t}}^2}{\bar{N}^2 \mu_f^2},M_{t\bar{t}},m_t,\cos\theta,\mu_f \right)\Bigg]
+ \mathcal{O}\left(\frac{1}{N}\right) \, ,
\end{multline}
for the hard-scattering kernel where the massive hard function $\bm{H}^{m}_{ij}$ is available up to NLO \cite{Ahrens:2010zv}, and the massive soft function $\tilde{\bm{s}}^m_{ij}$ has been calculated up to NNLO \cite{Wang:2018vgu}, though we use only the NLO result for consistency with the hard function in this work. Together with the anomalous dimensions governing the renormalization group evolution of these two functions,  resummation has been performed at NNLL accuracy in \cite{Ahrens:2010zv}. 

Besides soft logarithms, in the high energy regime when $M_{t\bar{t}} \gg m_t$, small-mass logarithms $\ln(m_t/M_{t\bar{t}})$ may also become important. In \cite{Ferroglia:2012ku}, a framework to simultaneously resum soft and small-mass logarithms was developed. The factorization formula for the hard-scattering kernel needed to achieve this resummation is given by\footnote{For simplicity, we have ignored the $n_h$-dependent contributions from top quark loops.}
\begin{multline}
\label{eq:boostedFac}
\tilde{c}_{ij}(N,\mu_f) = \Tr \left[ \bm{H}_{ij}(M_{t\bar{t}},\cos\theta,\mu_f) \, \tilde{\bm{s}}_{ij} \biggl(\ln\frac{M_{t\bar{t}}^2}{\bar{N}^2\mu_f^2},M_{t\bar{t}},\cos\theta,\mu_f\biggr) \right]
\\
\times C_D^2(m_t,\mu_f) \, \tilde{s}_D^2 \biggl(\ln\frac{m_t}{\bar{N}\mu_f},\mu_f\biggr) + \mathcal{O} \left( \frac{1}{N} \right) + \mathcal{O} \left( \frac{m_t}{M_{t\bar{t}}} \right) \, ,
\end{multline}
where the massless hard function $\bm{H}_{ij}$ and the massless soft function $\tilde{\bm{s}}_{ij}$ are defined in the limit $m_t \to 0$, and are available up to NNLO \cite{Broggio:2014hoa, Ferroglia:2012uy}. The logarithms of $m_t$ are absorbed into the two functions $C_D$ and $\tilde{s}_D$, which are related to collinear and soft-collinear emissions from the final state top quarks and are also available up to NNLO \cite{Ferroglia:2012ku}.\footnote{Note that the coefficient functions $C_D$, $\tilde{s}_D$ and $\tilde{\bm{s}}_{ij}$ used in this work are slightly different to those obtained in \cite{Ferroglia:2012ku, Ferroglia:2012uy}. We explain the details in appendix~\ref{sec:Revised}.} These ingredients allow resummation in this ``boosted-soft'' limit to be performed at NNLL$'$ order.

To obtain the resummed hard-scattering kernels, one has to derive and solve the RG equations for each of the matching functions in eqs.~(\ref{eq:soft-fact}) and~(\ref{eq:boostedFac}). This allows each function to be evaluated at a scale which frees it of large logarithms, and resums them in an evolution kernel. We denote the scales at which to evaluate the matching functions $\bm{H}^{(m)}_{ij}$, $\tilde{\bm{s}}^{(m)}_{ij}$, $C_D$, and $\tilde{s}_D$ by $\mu_h$, $\mu_s$, $\mu_{dh}$, and $\mu_{ds}$ respectively. Given the resummed hard-scattering kernels, a remaining difficulty in evaluating the formula (\ref{eq:x-sec2}) for the differential cross section is that the Mellin-space PDFs $\tilde{f}_{i/p}(N,\mu_f)$ are not readily available from program packages such as \texttt{LHAPDF} \cite{Buckley:2014ana}. It is possible to perform the Mellin-inversion on the kernel $\tilde{c}_{ij}(N,\mu_f)$, and carry out the convolution with the $x$-space PDFs as in, e.g. \cite{Catani:1996yz, Kulesza:2002rh}. However, this method suffers from numerical instabilities due to the fact that the resummed kernel $C_{ij}(z,\mu_f)$ is ill-behaved around the singular point $z = 1$. An alternative method described in \cite{Bonvini:2012sh} amounts to approximating the parton luminosity function
\begin{align}
\tilde{\mathcal{L}}_{ij}(N,\mu_f) \equiv \tilde{f}_{i/p}(N,\mu_f) \, \tilde{f}_{j/p}(N,\mu_f)
\end{align}
by an analytic expression with fitted coefficients. Briefly speaking, one can approximate the $x$-space version of the parton luminosity function
\begin{equation}
\mathcal{L}_{ij}(\xi,\mu_f) \equiv \int dx_1dx_2 \, \delta(\xi-x_1x_2) \, f_{i/p}(x_1,\mu_f) \, f_{j/p}(x_2,\mu_f) 
\end{equation}
by a linear combination of Chebyshev polynomials with $\mu_f$-dependent coefficients. The above formula can be numerically evaluated using inputs from \texttt{LHAPDF} and the coefficients can then be fitted using standard methods.  The Mellin transform of the Chebyshev polynomials can then be calculated analytically, which finally leads to an expression for the $N$-space luminosity function $\tilde{\mathcal{L}}_{ij}(N,\mu_f)$. Note that the fit has to be done for each distinct value of $\mu_f$, which leads to computational overheads if $\mu_f$ depends on the integration variables.

In \cite{Pecjak:2016nee, Czakon:2018nun}, we combined the above treatment of the parton luminosity function with the resummation framework in \cite{Ahrens:2010zv, Ferroglia:2012ku} to produce phenomenological predictions for the $t\bar{t}$ invariant mass distribution as well as the top-quark transverse momentum ($p_T$) distribution. The transverse momentum of the top quark is obtained in the soft limit via the relation $p_T = M_{t\bar{t}}\beta_t\sin\theta/2$. Since this relation does not involve the variables $x_1$, $x_2$ and $z$ in eq.~(\ref{eq:x-sec1}), the same luminosity function $\tilde{\mathcal{L}}_{ij}(N,\mu_f)$ can be used for a given $p_T$ phase-space point without difficulty. However, to extend our framework to rapidity distributions, we need to slightly modify the treatment of the PDFs, which we discuss in the following.

We will be concerned with two kinds of rapidities: the rapidity of the $t\bar{t}$ pair $Y_{t\bar{t}}$, and the rapidity of the top quark or the anti-top quark $y_{t/\bar{t}}$. In the soft limit we can express these as
\begin{align}
Y_{t\bar{t}} = \frac{1}{2} \ln \frac{x_1}{x_2} \, , \quad \hat{y} = \frac{1}{2} \ln \frac{1+\beta_t\cos\theta}{1-\beta_t\cos\theta} \, , \quad y_{t/\bar{t}} = Y_{t\bar{t}} \pm \hat{y} \, ,
\label{eq:rapidities}
\end{align}
where $\hat{y}$ is the rapidity of the top quark in the partonic center-of-mass frame. We start by rewriting eq.~(\ref{eq:x-sec1}) as
\begin{multline}
\frac{d^3\sigma(\tau)}{dM_{t\bar{t}}\,d\cos\theta\,dY_{t\bar{t}}} = \frac{8\pi\beta_t}{3sM_{t\bar{t}}}\sum\limits_{ij} \int dz d\xi dx_1 dx_2 \, \delta(\tau-z\xi) \, \delta(\xi - x_1x_2) \, \delta\bigg(Y_{t\bar{t}}-\frac{1}{2}\ln\frac{x_1}{x_2}\bigg)
\\
\times f_{i/p}(x_1,\mu_f) \, f_{j/p}(x_2,\mu_f) \, C_{ij}(z,\mu_f) \, ,
\end{multline}
and use the last two delta functions to integrate over $x_1$ and $x_2$ to arrive at
\begin{multline}
\frac{d^3\sigma(\tau)}{dM_{t\bar{t}}\,d\cos\theta\,dY_{t\bar{t}}} = \frac{8\pi\beta_t}{3sM_{t\bar{t}}}\sum\limits_{ij} \int \! dz \, d\xi \, \delta(\tau-z\xi) \,  C_{ij}(z,\mu_f)
\\
\times f_{i/p}\Big(\sqrt{\xi}e^{Y_{t\bar{t}}},\mu_f\Big) \, f_{j/p}\Big(\sqrt{\xi}e^{-Y_{t\bar{t}}},\mu_f\Big) \, .
\label{eq:x-sec3}
\end{multline}
Using techniques employed previously for the Drell-Yan process \cite{Bonvini:2010tp}, we define a rapidity-dependent parton luminosity function as
\begin{align}
L_{ij}(\xi,Y_{t\bar{t}},\mu_f) \equiv f_{i/p}\Big(\sqrt{\xi}e^{Y_{t\bar{t}}},\mu_f\Big) \, f_{j/p}\Big(\sqrt{\xi}e^{-Y_{t\bar{t}}},\mu_f\Big) \, ,
\end{align}
and its Mellin transform\footnote{This function is called $\mathscr{L}^{\text{rap}}(N,
\frac{1}{2})$ in \cite{Bonvini:2012sh}.}
\begin{align}
\tilde{L}_{ij}(N,Y_{t\bar{t}},\mu_f) \equiv \int_0^1 d\xi \,\xi^{N-1}\, L_{ij}(\xi,Y_{t\bar{t}},\mu_f) \, .
\end{align}
We can now perform a Mellin transform on eq.~(\ref{eq:x-sec3}) to arrive at
\begin{align}
\frac{d^3\tilde{\sigma}(N)}{dM_{t\bar{t}}\,d\cos\theta\,dY_{t\bar{t}}} &= \frac{8\pi\beta_t}{3sM_{t\bar{t}}}\sum\limits_{ij} \tilde{L}_{ij}(N,Y_{t\bar{t}},\mu_f) \, \tilde{c}_{ij}(N,\mu_f) \, .
\label{eq:x-sec4}
\end{align}
The new luminosity function $\tilde{L}_{ij}(N,Y_{t\bar{t}},\mu_f)$ can be approximated by an analytic expression using the same techniques as before. The formula (\ref{eq:x-sec4}) can be used to calculate single-variable differential cross sections, and can also be used to calculate double differential cross sections where two kinematic variables are measured simultaneously (see, e.g., \cite{Sirunyan:2018wem}). In this paper, we will only study the rapidity distributions, and leave the double differential cross sections to future work. 

It is straightforward to obtain the rapidity distribution of the $t\bar{t}$ pair by integrating over $M_{t\bar{t}}$ and $\cos\theta$ from eq.~(\ref{eq:x-sec4}). The integration ranges are given by
\begin{align}
-1 \le \cos\theta \le 1 \, , \quad 2m_t \le M_{t\bar{t}} \le \sqrt{s}/\cosh(Y_{t\bar{t}}) \, .
\end{align}
It is also easy to obtain the rapidity distribution of the top quark via a change of variables, which leads to
\begin{align}
\frac{d^3\tilde{\sigma}(N)}{dM_{t\bar{t}}\,d\cos\theta\,dy_t} &= \frac{8\pi\beta_t}{3sM_{t\bar{t}}}\sum\limits_{ij} \tilde{L}_{ij}(N,y_t-\hat{y},\mu_f) \, \tilde{c}_{ij}(N,\mu_f) \, ,
\label{eq:x-sec5}
\end{align}
where on the right-hand side, $\hat{y}$ should be expressed as a function of $M_{t\bar{t}}$ and $\cos\theta$ through eq.~(\ref{eq:rapidities}). A similar change of variables can be used to calculate the rapidity distribution of the anti-top quark. Phenomenologically, one is often interested in the average of the $y_t$ distribution and the $y_{\bar{t}}$ distribution, i.e.
\begin{align}
\label{eq:y_avt}
\frac{d\sigma}{dy_{\text{avt}}} \equiv \frac{1}{2} \left( \frac{d\sigma}{dy_t} \bigg|_{y_t=y_{\text{avt}}} + \frac{d\sigma}{dy_{\bar{t}}} \bigg|_{y_{\bar{t}}=y_{\text{avt}}} \right) \, .
\end{align}
In the next section, we will present our predictions for the $Y_{t\bar{t}}$ distribution and the $y_{\text{avt}}$ distribution.

Finally, to obtain precision predictions, it is necessary to combine the resummed results with fixed order ones whenever possible. The fixed order part accounts for formally power-suppressed terms, which can be important numerically. This ``matching'' procedure was described in detail in \cite{Czakon:2018nun}, and we refer the interested reader to that article.

\section{Numerical results}
\label{sec:results}

We now use the formalism introduced in the last section to produce numerical results relevant for LHC experiments at $\sqrt{s}=\unit{13}{\TeV}$. Throughout this section we set the top quark mass to $m_t=\unit{173.3}{\GeV}$. For both the $Y_{t\bar{t}}$ distribution and the $y_{\text{avt}}$ distribution, the default factorization scale is chosen to be $\mu_f^{\text{def}}=H_T/4$ following \cite{Czakon:2016dgf}, while the default values of the other matching scales are set as $\mu_h^{\text{def}}=H_T/2$, $\mu_s^{\text{def}}=H_T/\bar{N}$, $\mu_{dh}^{\text{def}}=m_t$ and $\mu_{ds}^{\text{def}}=m_t/\bar{N}$ following \cite{Czakon:2018nun}. Here $\bar{N} \equiv N e^{\gamma_E}$ with $\gamma_E$ denoting Euler's constant and $H_T$ is defined by
\begin{align}
H_T \equiv \sqrt{p_{T,t}^2+m_t^2} + \sqrt{p_{T,\bar{t}}^2+m_t^2} \, .
\end{align} 
We compare predictions using two different PDF sets:  CT14 \cite{Dulat:2015mca} and NNPDF3.1 \cite{Ball:2017nwa}, obtained from \texttt{LHAPDF} with ${\alpha_s(m_Z)=0.118}$. We match our resummed predictions to fixed-order results at NLO and NNLO. The NLO distributions \cite{Nason:1989zy, Mangano:1991jk, Frixione:1995fj} are generated using \texttt{MCFM} \cite{Campbell:2010ff}, while the NNLO ones \cite{Czakon:2016dgf} are obtained from \texttt{fastNLO} \cite{Kluge:2006xs, Wobisch:2011ij} using the tables available with \cite{Czakon:2017dip}. Note that the fastNLO tables do not provide scale variations and as a result, we only provide the central values for the NNLO(+NNLL$'$) predictions. On the other hand, for the NLO+NNLL$'$ predictions, we estimate the perturbative uncertainties by varying each of the scales individually up and down by a factor of two and combining the resulting variations of the differential cross sections in quadrature. All our predictions are calculated using NNLO PDFs. Only when we show the NLO predictions for comparison do we use NLO PDFs.

\begin{figure}[t!]
\centering
\includegraphics[width=0.495\textwidth]{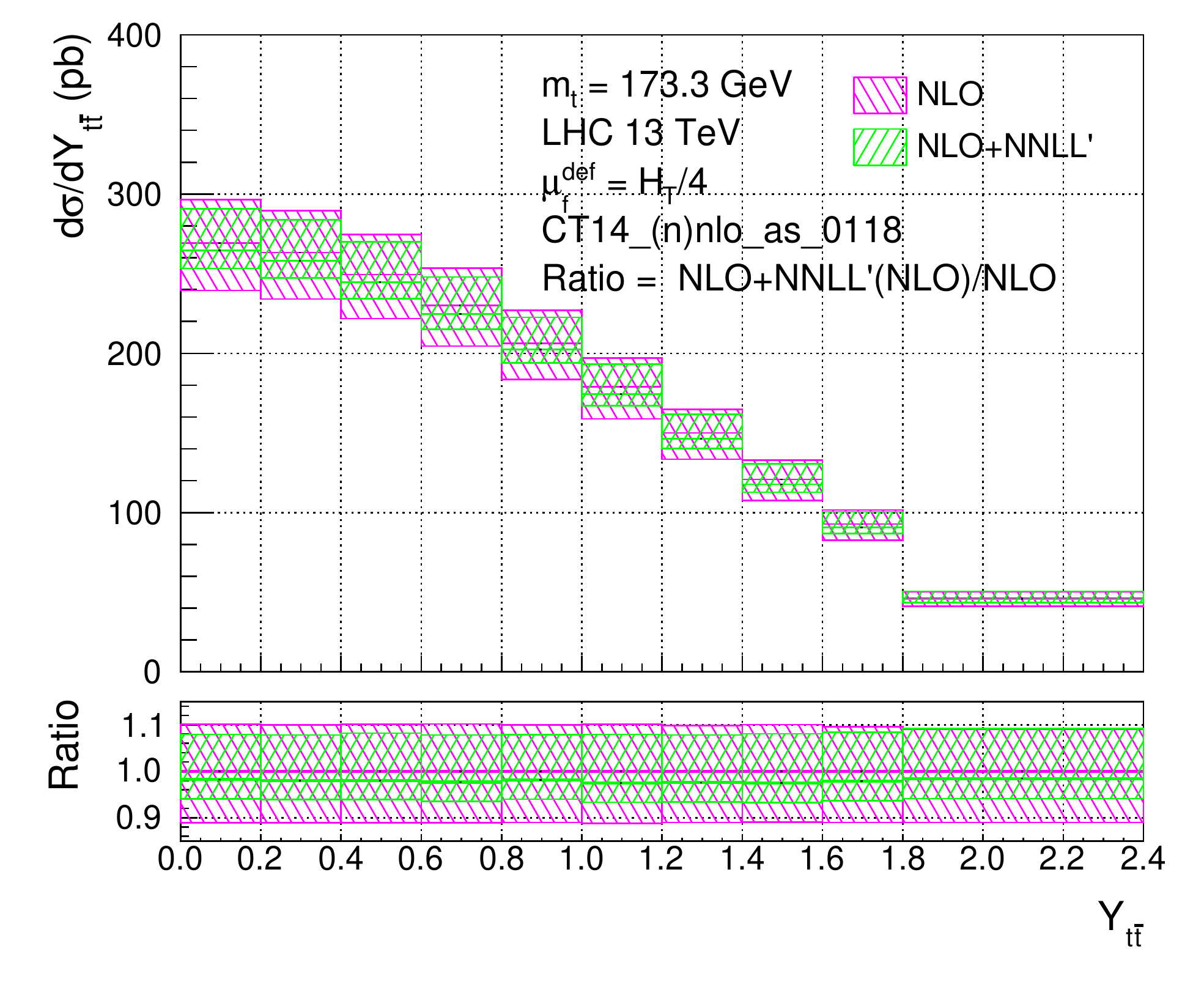}
\includegraphics[width=0.495\textwidth]{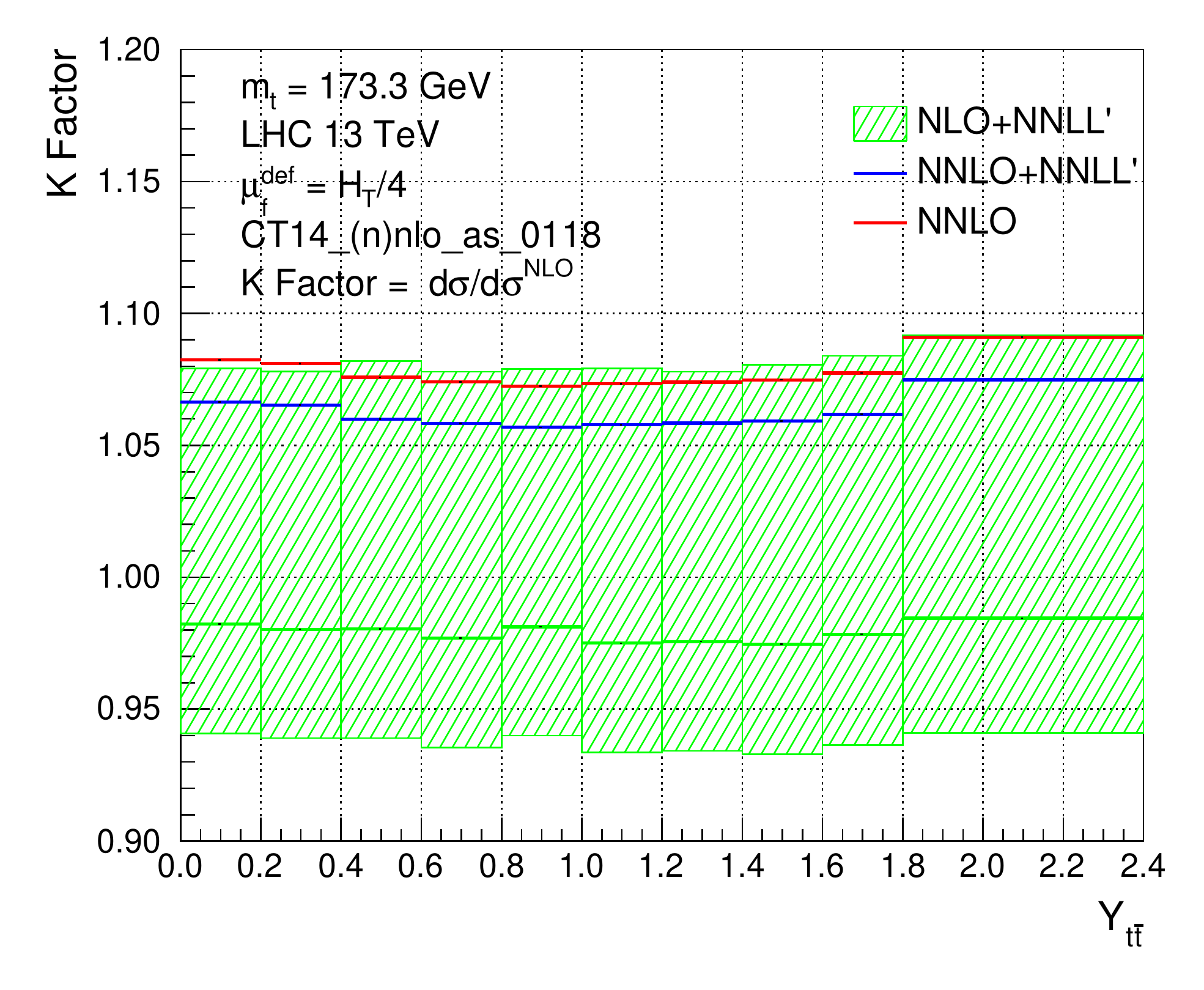}
\vspace{-6ex}
\caption{\label{fig:ytt1}Left: the $Y_{t\bar{t}}$ distribution at NLO and NLO+NNLL$'$ with scale uncertainties; Right: the ratios (K-factors) of NNLO and NNLO+NNLL$'$ to NLO with central values only, and NLO+NNLL$'$ with scale uncertainties. Results are obtained using the CT14 PDF sets.}
\end{figure}

In figure~\ref{fig:ytt1} we compare our resummed predictions for the $Y_{t\bar{t}}$ distribution to the fixed-order results at NLO and NNLO. We show the results with the CT14 PDF sets here. The left plot shows a comparison between the NLO and NLO+NNLL$'$ predictions, where in the lower panel we show the ratio defined by
\begin{equation}
\text{Ratio}=\frac{d\sigma}{d\sigma^{\text{NLO}}(\mu_f^{\text{def}})} \, .
\end{equation}
We see that the two results are rather similar, with the scale dependence slightly reduced by the resummation. On the right plot of figure~\ref{fig:ytt1}, we compare the central values of the NNLO and the NNLO+NNLL$'$ predictions by displaying their ratios to the NLO result (the K-factors). The NLO+NNLL$'$ result is also shown (with uncertainty) as a reference. We find that matching our resummation to NNLO compared to NLO predictions increases the central value by about 8\%. We also find that the central values of the NNLO and NNLO+NNLL$'$ results are close to each other. This implies that with the choice $\mu_f=H_T/4$ in fixed-order calculations, higher order corrections beyond NNLO are well under control,  consistent with the behavior of the $M_{t\bar{t}}$ and $p_T$ distributions studied in \cite{Czakon:2018nun}. 

\begin{figure}[t!]
\centering
\includegraphics[width=0.495\textwidth]{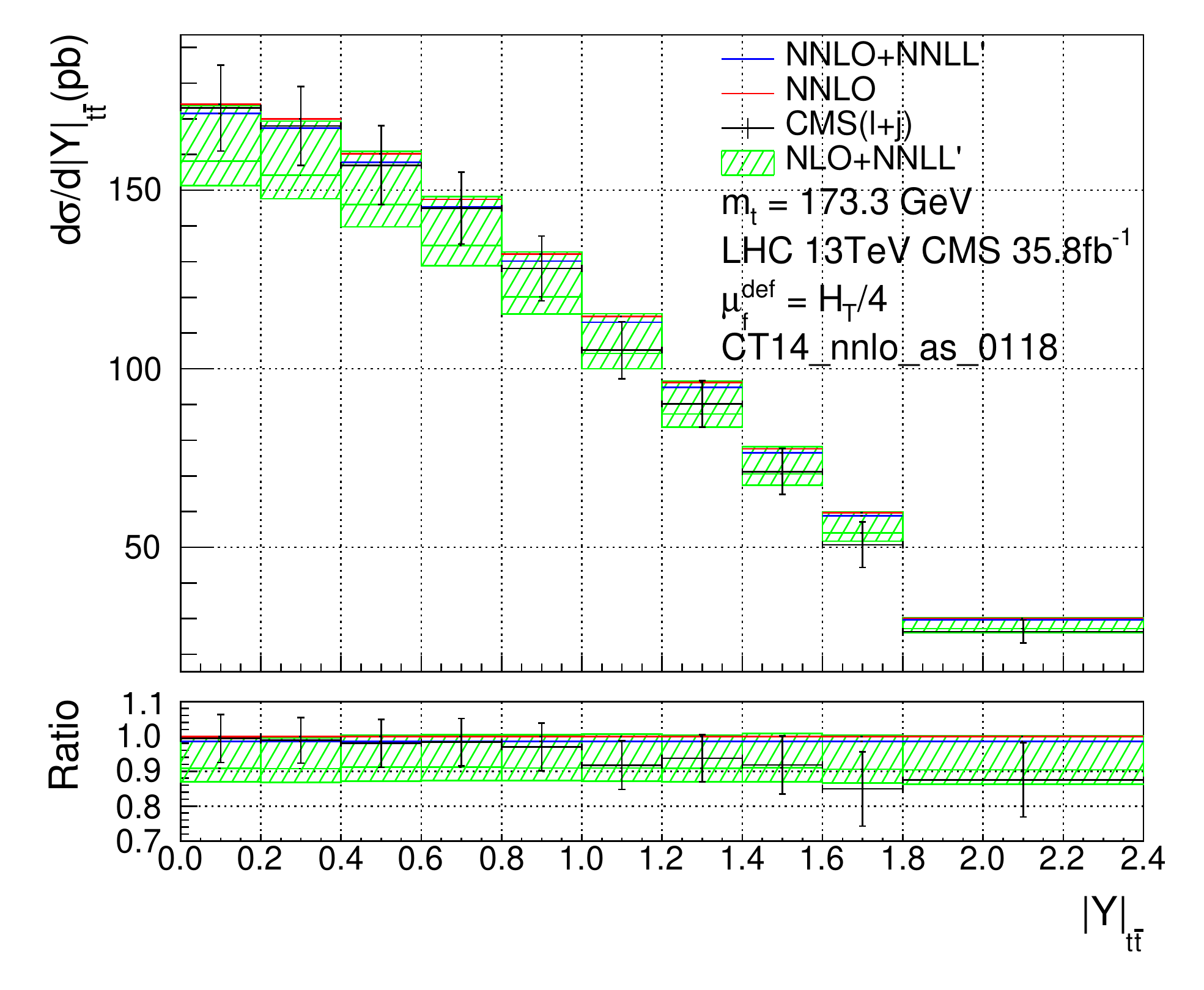}
\includegraphics[width=0.495\textwidth]{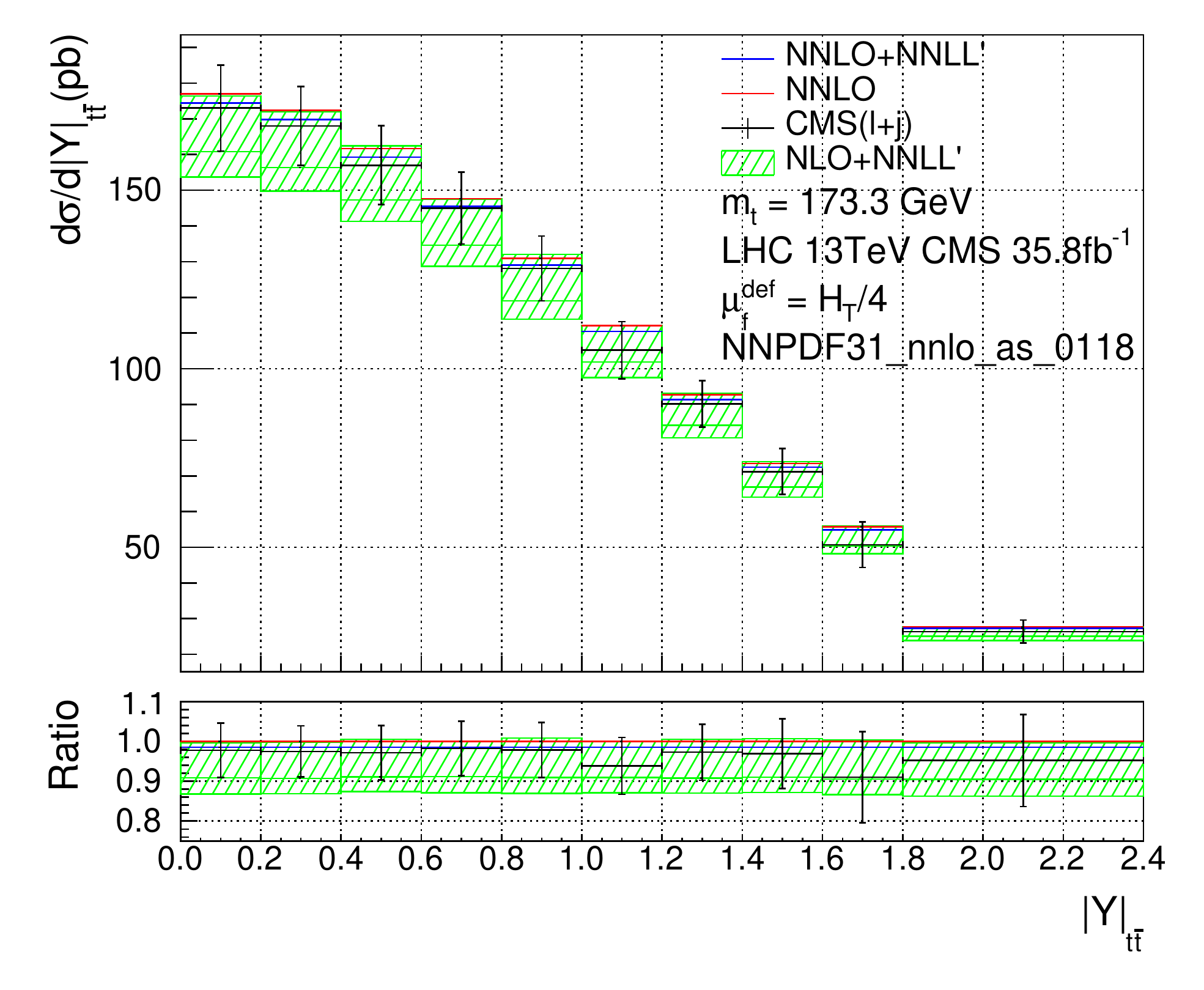}
\vspace{-6ex}
\caption{\label{fig:ytt2}Comparison between theoretical predictions and experimental data from the CMS collaboration \cite{Sirunyan:2018wem} for the $|Y_{t\bar{t}}|$ distribution in the lepton+jets channel. The left and right plots use CT14 and NNPDF3.1 PDF sets, respectively. In the lower panels we normalize everything to the NNLO central values.}
\end{figure}

In figure~\ref{fig:ytt2} we compare the theoretical predictions to recent experimental measurement from the CMS collaboration using $\unit{35.8}{\invfb}$ of data \cite{Sirunyan:2018wem}. The left and the right plots show predictions produced with the CT14 and the NNPDF3.1 PDF sets, respectively. The CMS results are given in terms of the differential production rate in the lepton+jets channel. In order to compare with that, we must rescale our results by the branching ratio of the $t\bar{t}$ pair decaying into the $b\bar{b}l\nu j j$ final state, where $l=e$, $\mu$. We assume the top quarks decay exclusively to a bottom quark and a $W$ boson, and extract the theoretical predictions for the braching ratios of the $W \to l\nu$ decay and the $W \to q\bar{q}'$ decay from the PDG review \cite{Tanabashi:2018oca}. The rescaling factor for our predictions is therefore given by
\begin{equation}
  \label{eq:br}
  \mathrm{Br}(t\bar{t} \to b\bar{b}l\nu j j) = 0.2985 \, .
\end{equation}
After the rescaling, the theoretical predictions are in excellent agreement with data. However, we find that the predictions from the two PDF sets have slightly different shapes. Especially in the tail region (large $|Y_{t\bar{t}}|$), the CT14 PDFs tend to predict a higher production rate than the NNPDF3.1 PDFs. It is possible to exploit this region to gain further information about the parton distributions, which we leave for future investigation.

\begin{figure}[t!]
\centering
\includegraphics[width=0.495\textwidth]{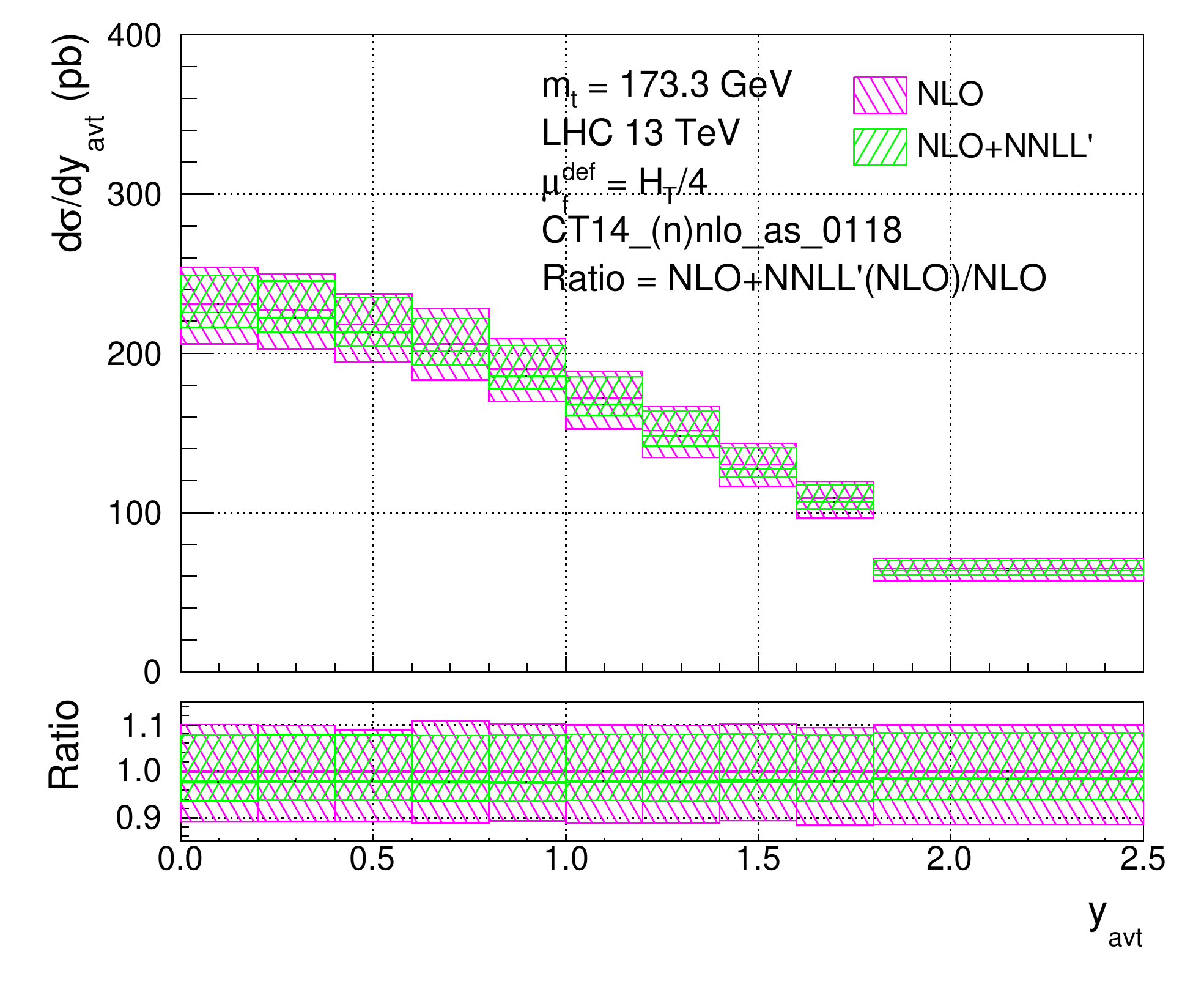}
\includegraphics[width=0.495\textwidth]{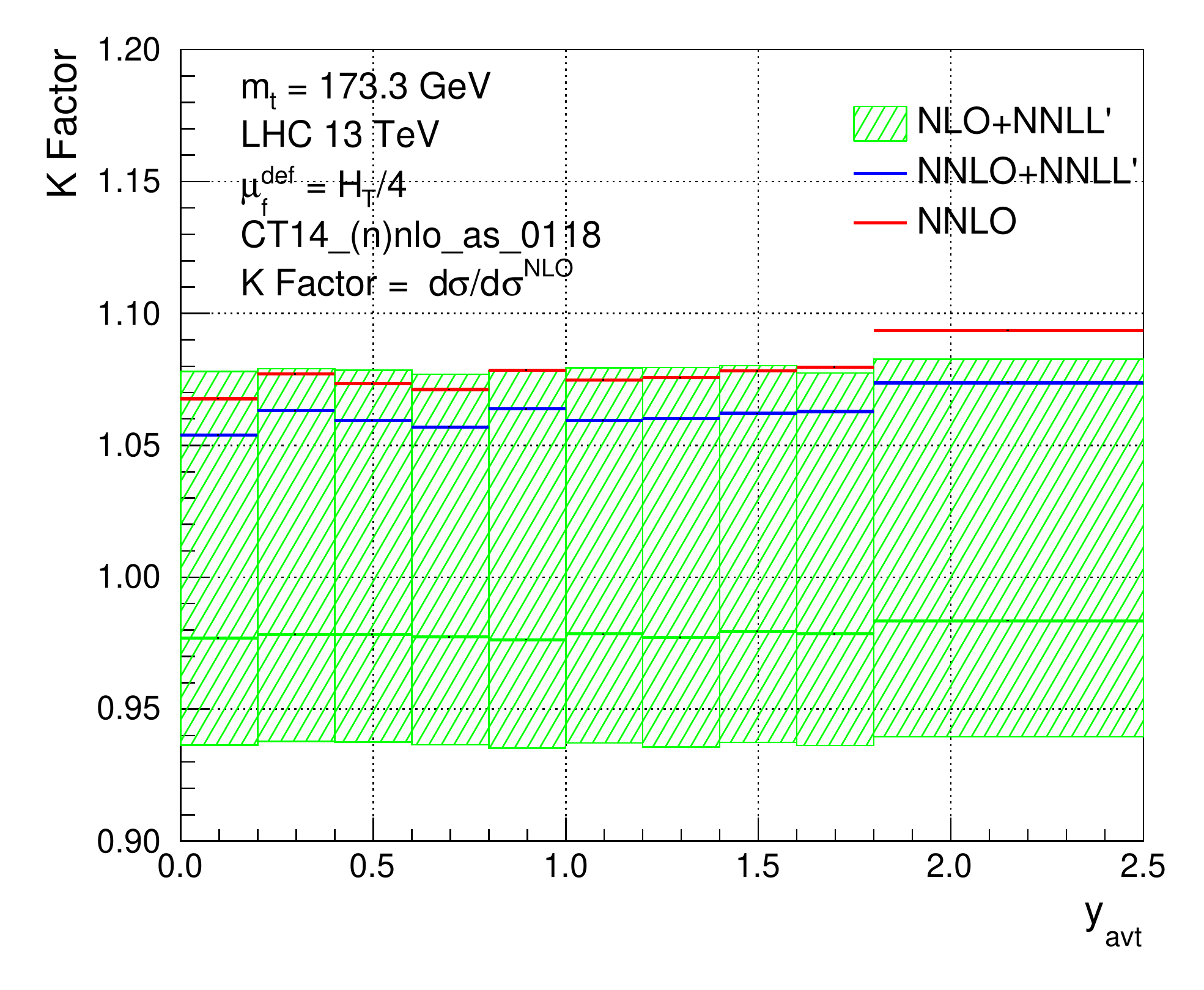}
\vspace{-6ex}
\caption{\label{fig:yt1}Left: the $y_{\text{avt}}$ distribution at NLO and NLO+NNLL$'$ with scale uncertainties; Right: the ratios (K-factors) of NNLO and NNLO+NNLL$'$ to NLO with central values only, and the NLO+NNLL$'$ with scale uncertainties. Results are obtained using the CT14 PDF sets.}
\end{figure}
\begin{figure}[t!]
\centering
\includegraphics[width=0.495\textwidth]{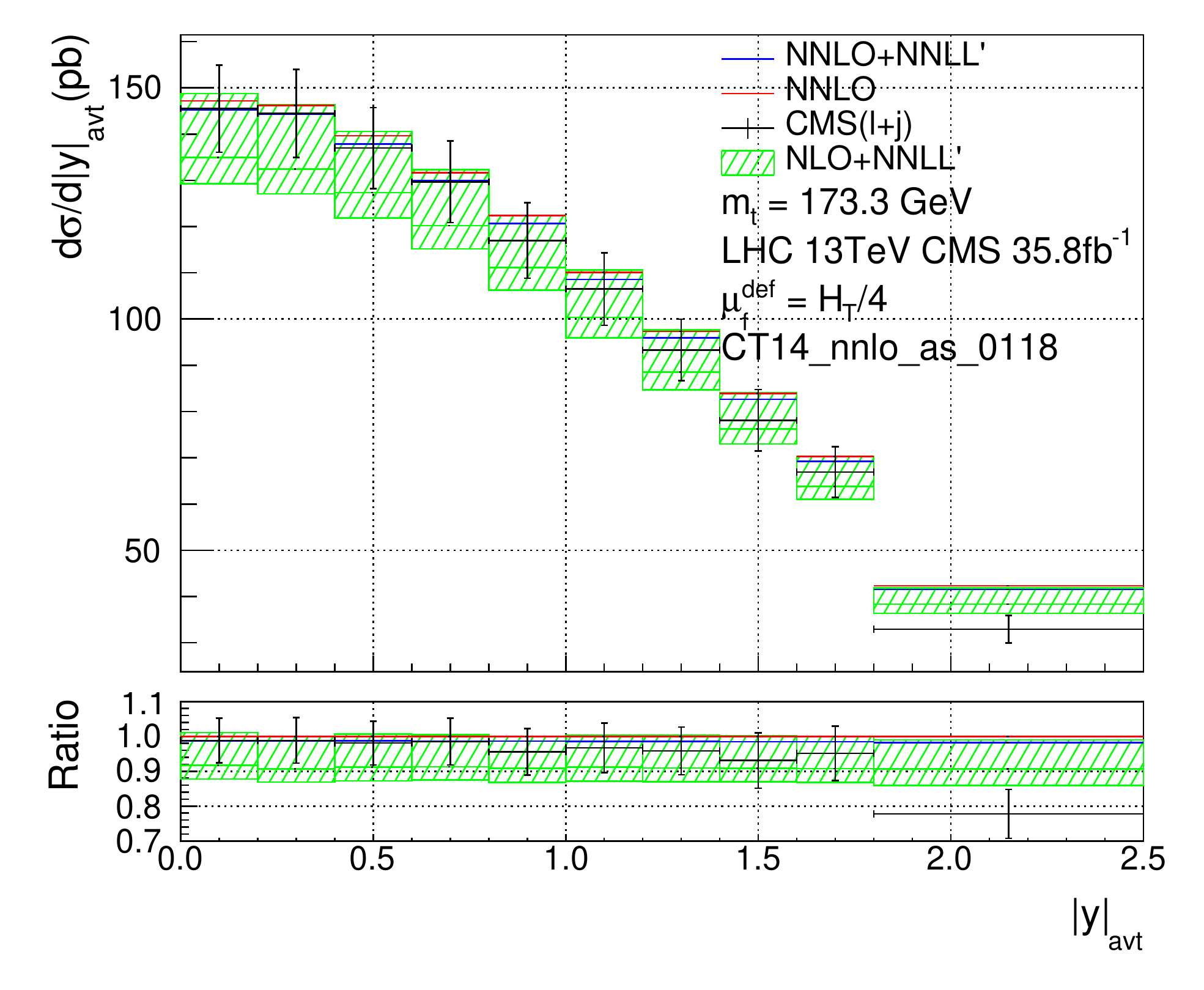}
\includegraphics[width=0.495\textwidth]{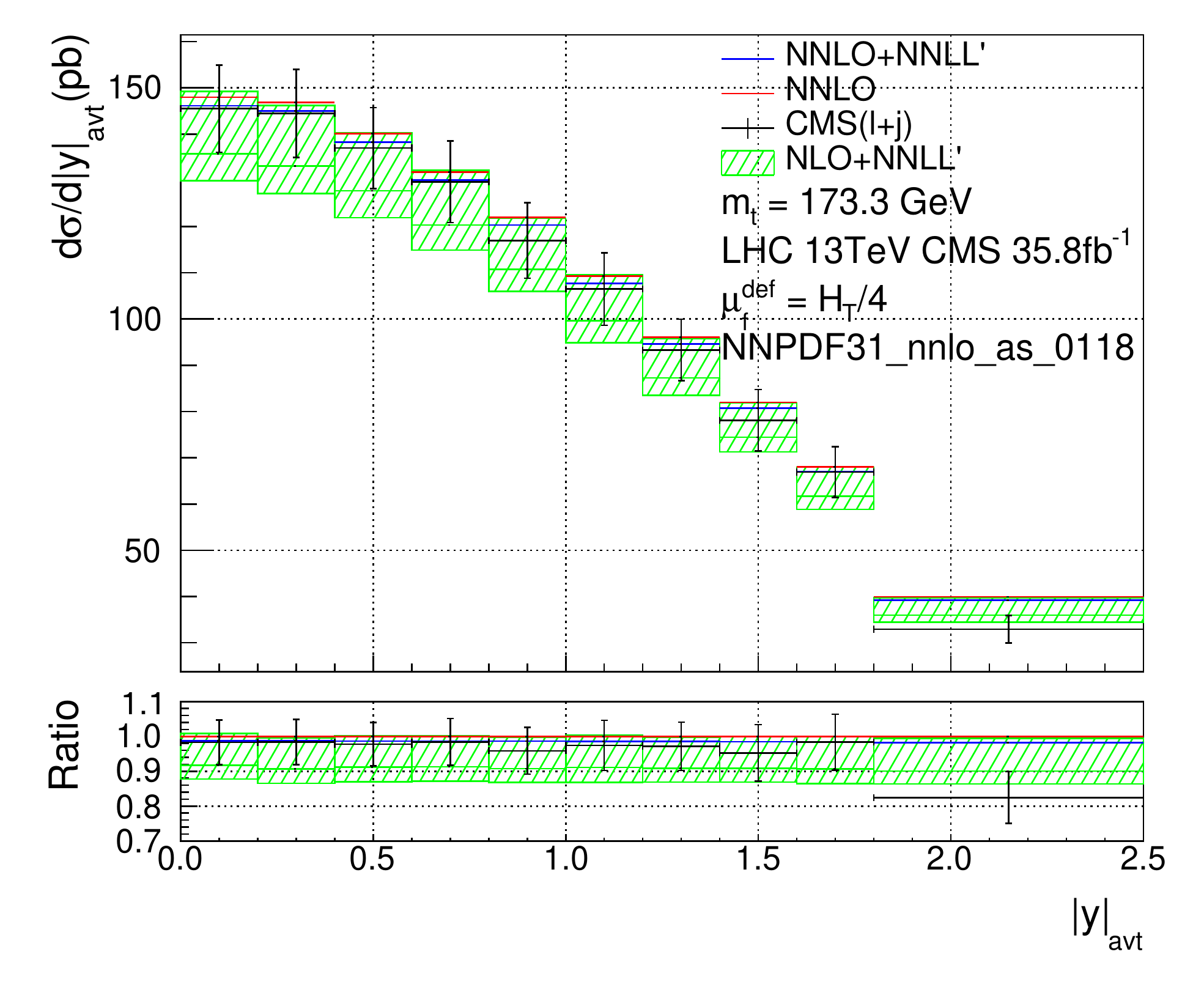}
\vspace{-6ex}
\caption{\label{fig:yt2}Comparison between theoretical predictions and the experimental data from the CMS collaboration \cite{Sirunyan:2018wem} for the $|y_{\text{avt}}|$ distribution in the lepton+jets channel. The left and right plots use CT14 and NNPDF3.1 PDF sets, respectively. In the lower panels we normalize everything to the NNLO central values.}
\end{figure}

We now turn to the average rapidity distribution defined in eq.~(\ref{eq:y_avt}). Figure~\ref{fig:yt1} shows predictions for the $y_{\text{avt}}$ distribution in a form analogous to figure~\ref{fig:ytt1}. We observe behavior similar to that of the $Y_{t\bar{t}}$ distribution: that the resummation effects reduce the perturbative uncertainties, and that the NNLO+NNLL$'$ results are close to the NNLO ones. We then compare the various theoretical predictions to the CMS data in figure~\ref{fig:yt2}. As before, we need to rescale the theoretical results by the factor in eq.~(\ref{eq:br}). We again find a good overall agreement except for the last bin, where the theoretical predictions tend to overestimate the cross section. We also see that the predictions from the two PDF sets are slightly different here, hinting that the $y_{\text{avt}}$ distribution may also be used to improve the PDF fitting.

\begin{figure}
  \centering
  \includegraphics[width=0.495\textwidth]{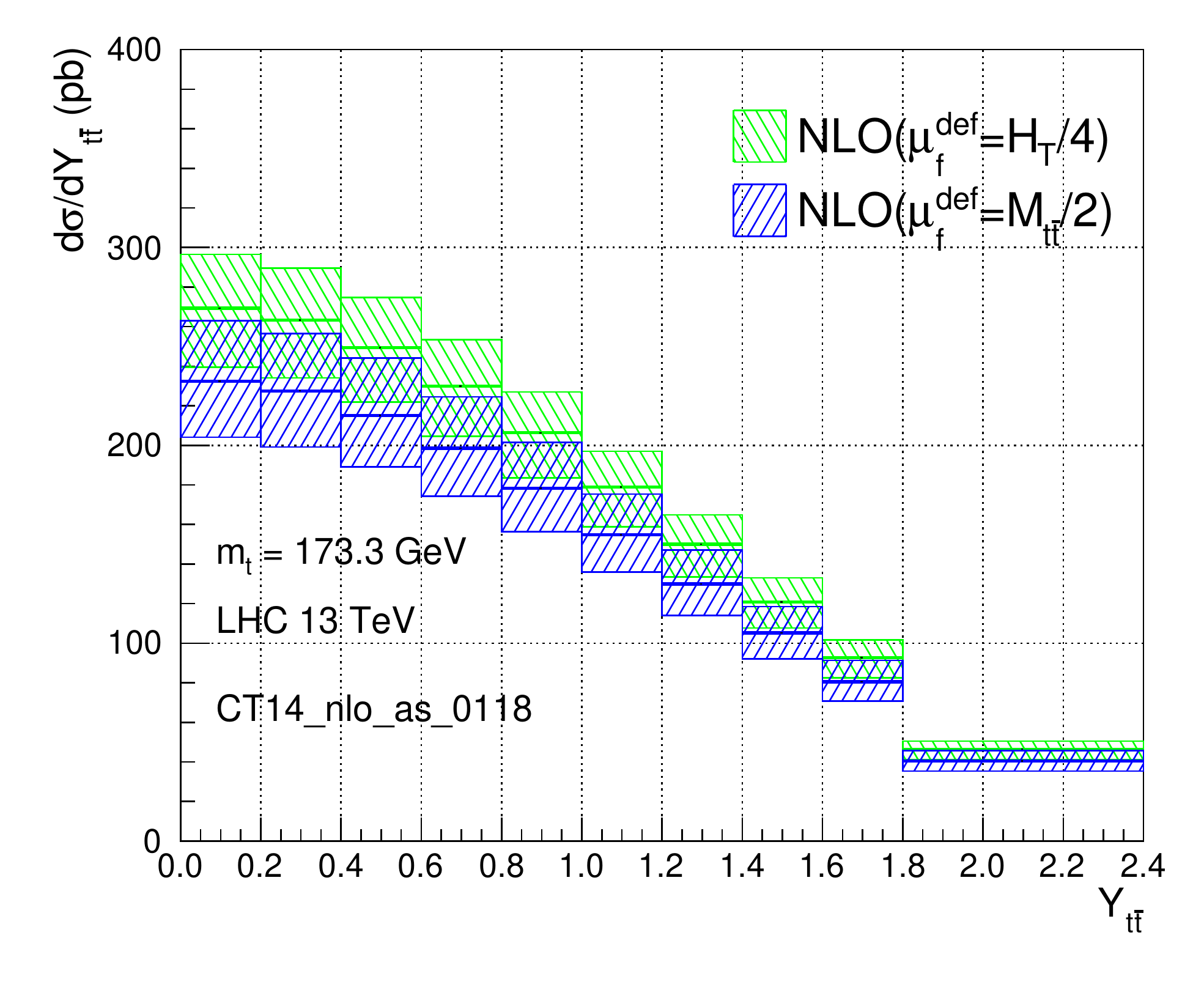}
  \includegraphics[width=0.495\textwidth]{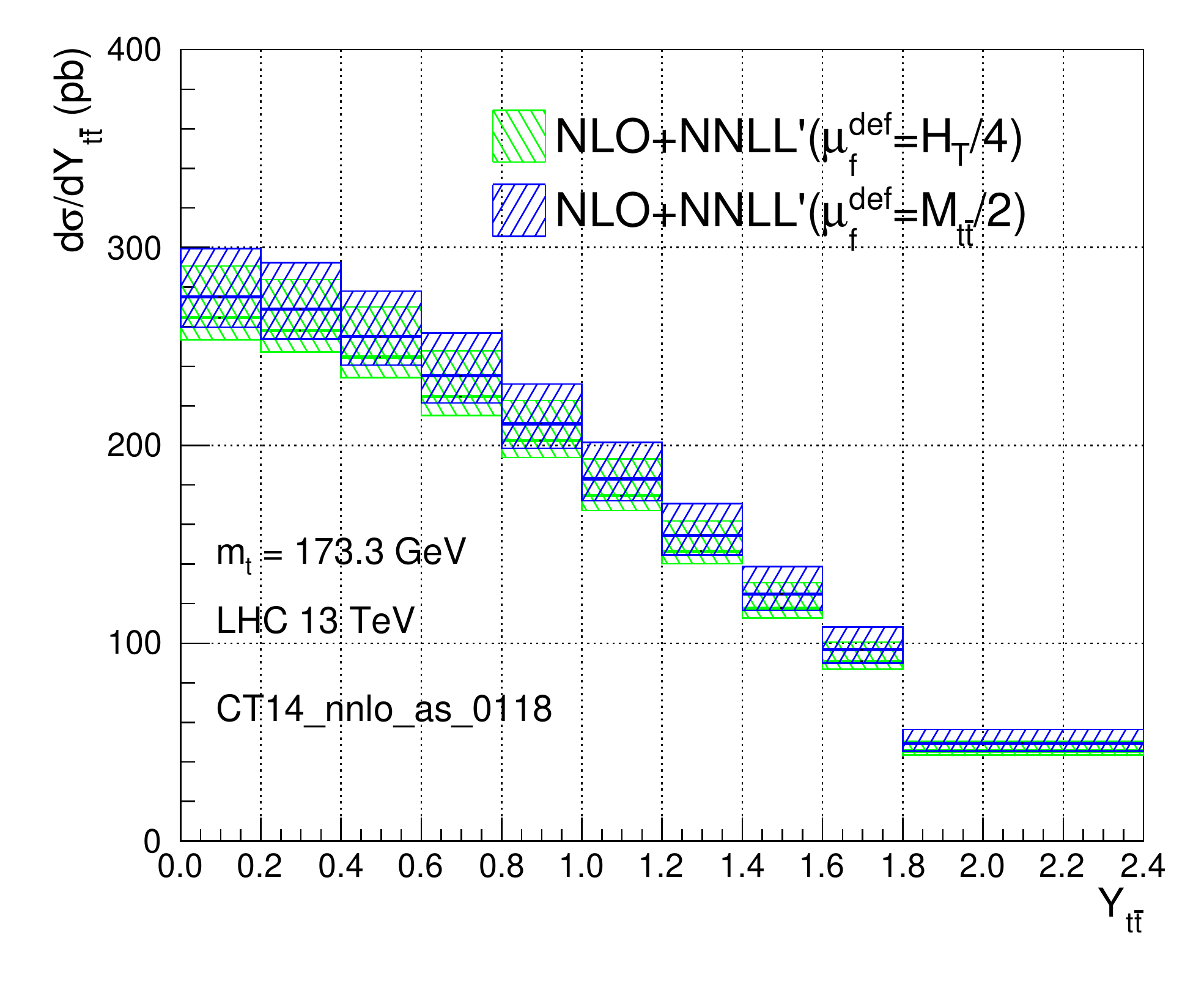}
  \\
  \includegraphics[width=0.495\textwidth]{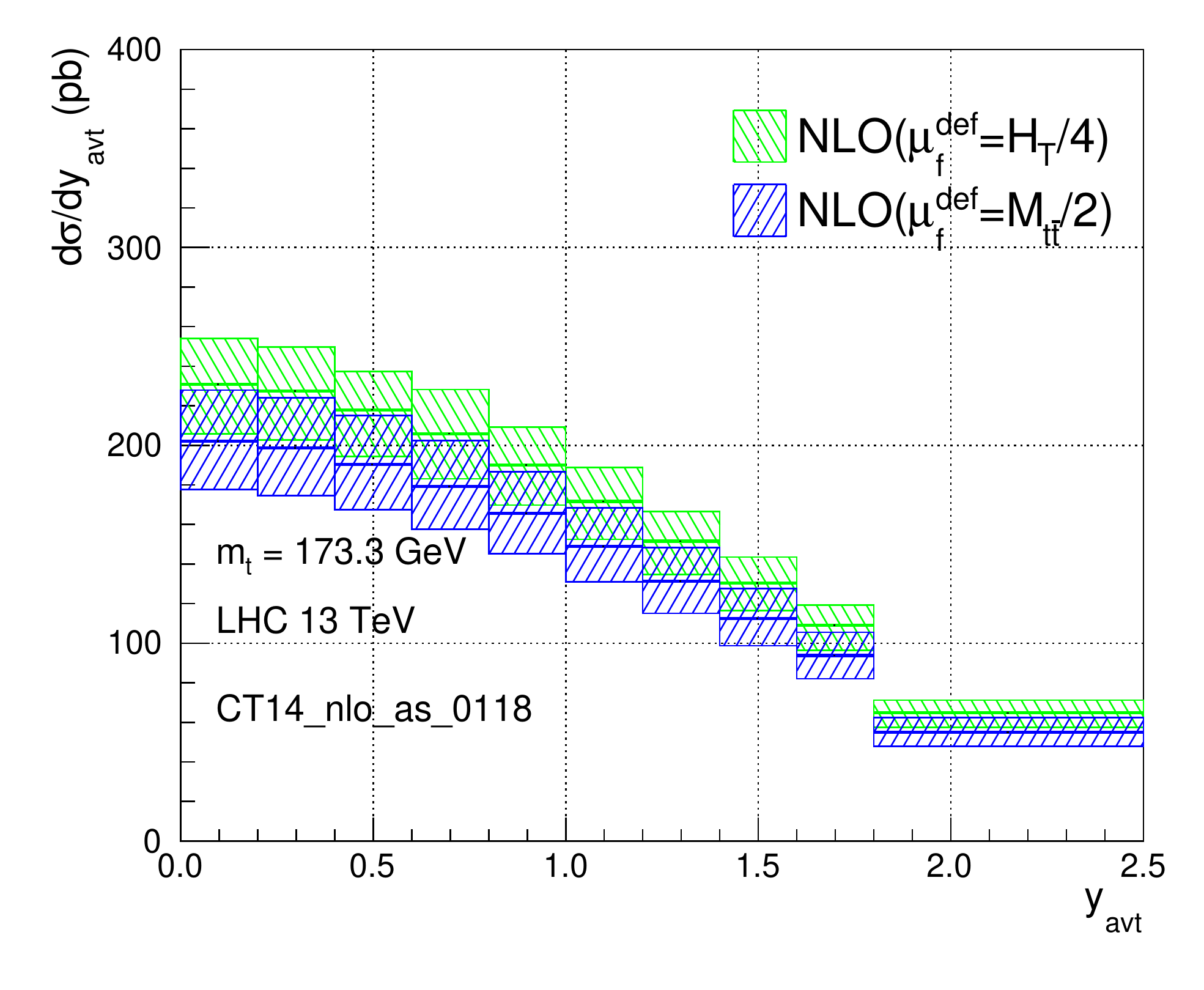}
  \includegraphics[width=0.495\textwidth]{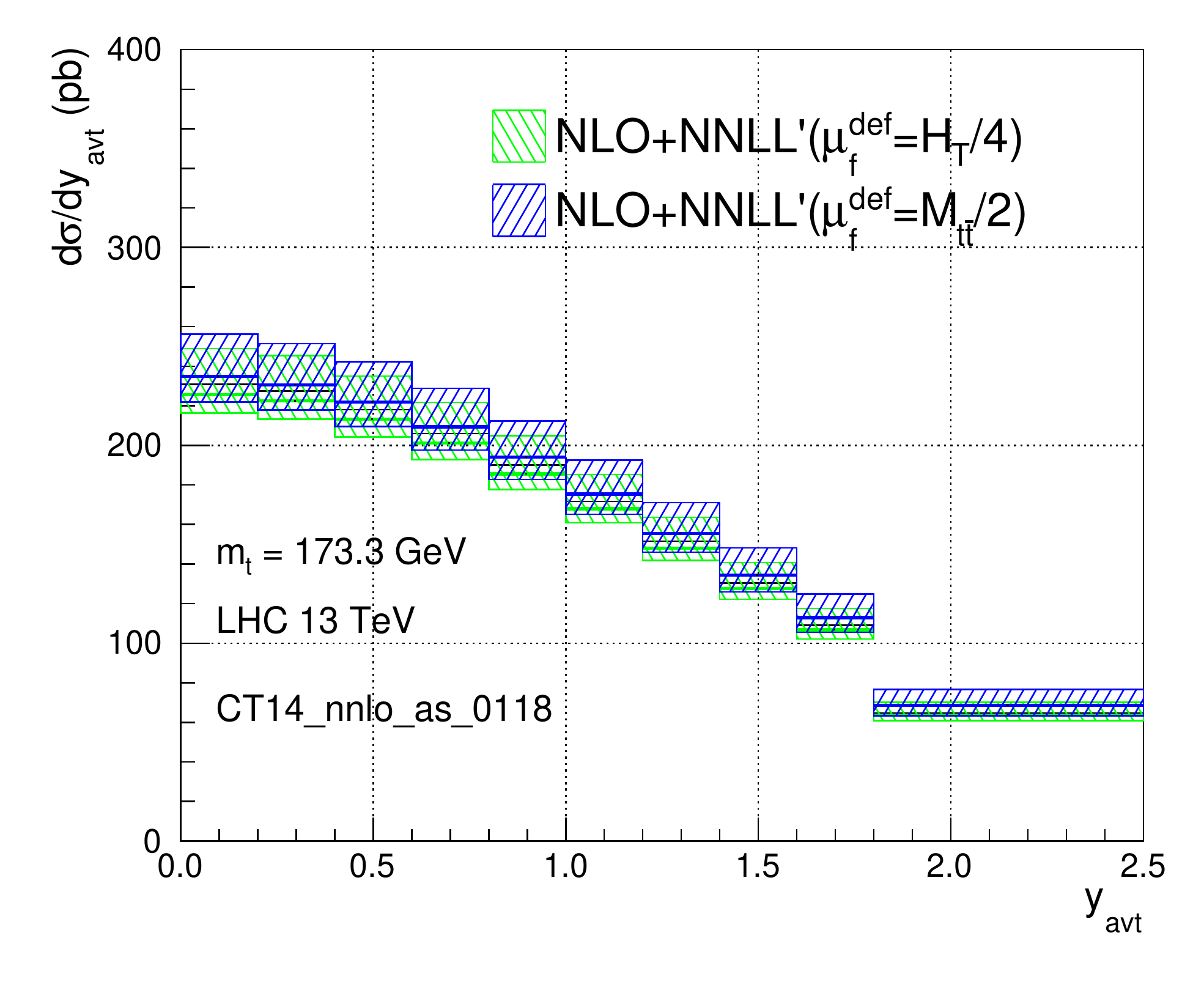}
  \vspace{-6ex}
  \caption{\label{fig:scale}Comparison between the two default choices for the factorization scale, $\mu_f^{\text{def}} = H_T/4$ and $\mu_f^{\text{def}}=M_{t\bar{t}}/2$.}
\end{figure}

One of the important findings of \cite{Czakon:2018nun} is that resummation effects stabilize the dependence of the differential cross sections on the choice of the factorization scale $\mu_f$. Given the new results in this work, it is interesting to do the same comparison for the rapidity distributions. In figure~\ref{fig:scale}, we present the NLO and the NLO+NNLL$'$ predictions with two default choices for $\mu_f$: $\mu_f^{\text{def}} = H_T/4$ and $\mu_f^{\text{def}}=M_{t\bar{t}}/2$, with the other scales chosen as before. We see that for both the $Y_{t\bar{t}}$ distribution and the $y_{\text{avt}}$ distribution, the NLO+NNLL$'$ results exhibit smaller sensitively to the choice of $\mu_f^{\text{def}}$ than the NLO ones. This is similar in spirit to the conclusion of \cite{Czakon:2018nun}.

Finally, it would be interesting to compare the scale dependence of the NNLO and the NNLO+NNLL$'$ results. We have all ingredients in our resummation formula ready for the NNLO+NNLL$'$ matching including scale variations, and it will be straightforward to combine them with the NNLO results from \cite{Czakon:2016dgf}.  Based on the experience gained from the studies of differential
distributions in \cite{Czakon:2018nun}, we would expect that the NNLO+NNLL$'$ rapidity distributions are
close to the NNLO ones for the choice $\mu_f^{\text{def}} = H_T/4$, and that they are less sensitive to the different choices of the default factorization scale. We hope to validate these expectations in the future.

\section{Conclusions}
\label{sec:conclusions}

In this work we have extended the resummation framework for top quark pair production to the rapidity distribution of the $t\bar{t}$ system ($Y_{t\bar{t}}$), as well as the rapidity distributions of the top quark and the anti-top quark (and their average $y_{\text{avt}}$). Predictions have been presented both at NLO+NNLL$'$ and NNLO+NNLL$'$ accuracy, and are compared to recent experimental measurements in the lepton + jets channel by the CMS collaboration with good agreement. We find that the resummation effects are mild with the default scale choice $\mu_f^{\text{def}}=H_T/4$, showing that higher order corrections beyond NNLO are under control. The fixed-order results are, however, more sensitive to the choice of $\mu_f$. When $\mu_f^{\text{def}}=M_{t\bar{t}}/2$, for example, the NLO predictions are much lower than the results obtained with $\mu_f^{\text{def}}=H_T/4$. In contrast, the NLO+NNLL$'$ results are affected much less by parametric changes of the default factorization scale.

In the comparison with experimental data, we find that the theoretical predictions in the boosted regions (high $|Y_{t\bar{t}}|$ or high $|y_{\text{avt}}|$) exhibit some differences when different PDF sets are used. This region can thus be used to constrain the gluon PDF in the future, which is indispensable for improving the precision of the theoretical prediction for Higgs boson production.

In appendix~\ref{sec:Revised}, we have presented some slight differences in the coefficient functions used in this article compared to those used in earlier works. We show that the modifications have minimal impact on the $M_{t\bar{t}}$ and $p_{T}$ distributions, and therefore the results of the earlier works are not altered.

\section*{Acknowledgements}

This work was supported in part by the National Natural Science Foundation of China under Grant No. 11575004 and 11635001. D.J.S.~is supported under the ERC grant ERC-STG2015-677323.
\appendix

\section{Modifications to the coefficient functions}
\label{sec:Revised}

The coefficients $C_D$ and $\widetilde{s}_D$  appearing in the factorization formula (\ref{eq:boostedFac}) were extracted in \cite{Ferroglia:2012ku} and subsequently used in \cite{Pecjak:2016nee, Czakon:2018nun}. However, it was noted in \cite{Ferroglia:2012ku} that there exists an inconsistency in the literature which may lead to a shift of a constant term at order $\alpha_s^2$ between $C_D$ and $\widetilde{s}_D$. This shift has indeed been confirmed by the explicit calculation of the NNLO massive soft function $\widetilde{\bm{s}}^m_{ij}$ in \cite{Wang:2018vgu}, for which the small-mass factorization can be validated by taking the limit $m_t \ll M_{t\bar{t}}$. Another by-product of the calculation in \cite{Wang:2018vgu} is the extraction of a purely-imaginary off-diagonal contribution to the massless soft function $\widetilde{\bm{s}}_{ij}$, which had previously been omitted in \cite{Ferroglia:2012uy}. This contribution arises from gluon exchanges among 3 Wilson lines, and is important for both the massive and massless soft functions to satisfy their RG equations.

For these reasons, the coefficient functions $\widetilde{\bm{s}}_{ij}$, $C_D$ and $\widetilde{s}_D$ used in this work differ slightly to those used in \cite{Pecjak:2016nee, Czakon:2018nun}. The purpose of this appendix is to present these modifications and to assess their numerical impact on the $M_{t\bar{t}}$ and $p_{T,\text{avt}}$ distributions.

The matching functions $C_D$ and $\widetilde{s}_D$ admit a perturbative expansion in $\alpha_s$ which we write as
\begin{align}
C_D &= 1 + \frac{\alpha_s}{4\pi} C_D^{(1)} + \left(\frac{\alpha_s}{4\pi}\right)^2 C_D^{(2)} + \cdots \, , \nonumber
\\
\widetilde{s}_D &= 1 + \frac{\alpha_s}{4\pi} \widetilde{s}_D^{(1)} + \left(\frac{\alpha_s}{4\pi}\right)^2 \widetilde{s}_D^{(2)} + \cdots \, .
\end{align}
The modification concerns the NNLO coefficients $C_D^{(2)}$ and $\widetilde{s}_D^{(2)}$. We denote the expressions of these coefficients used in \cite{Pecjak:2016nee, Czakon:2018nun} by $C_{D,\text{old}}^{(2)}$ and $\widetilde{s}_{D,\text{old}}^{(2)}$, while the correct ones used in this work by $C_{D,\text{new}}^{(2)}$ and $\widetilde{s}_{D,\text{new}}^{(2)}$. They are related by
\begin{equation}
\label{eq:cDsD_shift}
C_{D,\text{new}}^{(2)} = C_{D,\text{old}}^{(2)} - 4\pi^2C_AC_F\, , \qquad \widetilde{s}_{D,\text{new}}^{(2)}=\widetilde{s}_{D,\text{old}}^{(2)} + 4\pi^2C_AC_F \, .
\end{equation}

Another modification concerns the NNLO massless soft function $\widetilde{\bm{s}}^{(2)}_{ij}$. Again we denote the old and the new expressions with corresponding subscripts. They are related by
\begin{equation}
\widetilde{\bm{s}}^{(2)}_{ij,\text{new}} = \widetilde{\bm{s}}^{(2)}_{ij,\text{old}} + \widetilde{\bm{s}}^{(2)}_{ij,\text{3w}} \, .
\end{equation}
The subscript ``3w'' makes it clear that this contribution arises from correlations among 3 Wilson lines, which is non-vanishing in the virtual-real diagrams at NNLO. The soft function is a Hermitian matrix in color space, and therefore it is enough to give the non-zero entries in the upper-right part. For the $q\bar{q}$ channel, we have
\begin{multline}
  \left(\widetilde{\bm{s}}^{(2)}_{q\bar{q},\text{3w}}\right)_{12} = 32 i \pi \bigg[ \frac{L^2}{2} \big( H_0(x_t) +  H_1(x_t) \big) + L \big( H_{0,0}(x_t) + H_{1,0}(x_t) - H_{1,1}(x_t) - H_2(x_t) + \zeta_2 \big)
  \\
  + H_{1,2}(x_t) + H_{2,0}(x_t) + 2 H_{2,1}(x_t) + H_{0,0,0}(x_t) + 2 H_{1,0,0}(x_t)
  \\
  + 2 H_{1,1,0}(x_t) + H_{1,1,1}(x_t) + 2 H_3(x_t) + \frac{\pi^2}{3} H_0(x_t) + \frac{\pi^2}{2} H_1(x_t) - 2 \zeta_3 \bigg] \, ,
\end{multline}
where
\begin{equation}
x_t=(1-\beta_t\cos\theta)/2 \, , \quad L=\ln\left(\frac{M_{t\bar{t}}^2}{\bar{N}^2 \mu^2}\right) \, .
\end{equation}
The contribution for the $gg$ channel is given by
\begin{equation}
\left(\widetilde{\bm{s}}^{(2)}_{gg,\text{3w}}\right)_{12} = \frac{9}{4} \left(\widetilde{\bm{s}}^{(2)}_{q\bar{q},\text{3w}}\right)_{12} \, .
\end{equation}
In the above formulas we have set the number of colors $N_c = 3$ and the number of light quarks $N_l=5$. The $H_{i,j,..}$ functions denote harmonic polylogarithms with weights specified by the indices $\{i,j,..\}$.

\begin{table}[t!]
\centering
\begin{tabular}{|c|c|r|r|}
  \hline
  observable & bin & old
  & new
  \\ \hline
  \multirow{2}{*}{$M_{t\bar{t}}$} & \unit{[580,620]}{\GeV} & \unit{1.129}{\picobarn}
                                                    & \unit{1.132}{\picobarn}
  \\ \cline{2-4}
             & \unit{[2500,3000]}{\GeV} & \unit{$1.548\times10^{-4}$}{\picobarn}
                                                    & \unit{$1.547\times10^{-4}$}{\picobarn}
  \\ \hline
  \multirow{2}{*}{$p_{T,\text{avt}}$} & \unit{[50,100]}{\GeV} & \unit{4.830}{\picobarn}
                                                    & \unit{4.847}{\picobarn}
  \\ \cline{2-4} 
             & \unit{[800,900]}{\GeV} & \unit{$7.743\times10^{-4}$}{\picobarn}
                                                    & \unit{$7.789\times10^{-4}$}{\picobarn}
  \\ \hline                                                            
\end{tabular}
\caption{\label{tab:cDsD}Comparison of the numerical results computed with the old (used in \cite{Pecjak:2016nee, Czakon:2018nun}) and the new (used in this work) coefficient functions $\widetilde{\bm{s}}_{ij}$, $C_D$ and $\widetilde{s}_D$. We show the central values with the default scale choices given in \cite{Czakon:2018nun}. The results here are obtained using the NNPDF3.0 PDF sets with ${\alpha_s(M_Z)=0.118}$ \cite{Ball:2014uwa}.}
\end{table}

\begin{figure}[t!]
\centering
\includegraphics[width=0.495\textwidth]{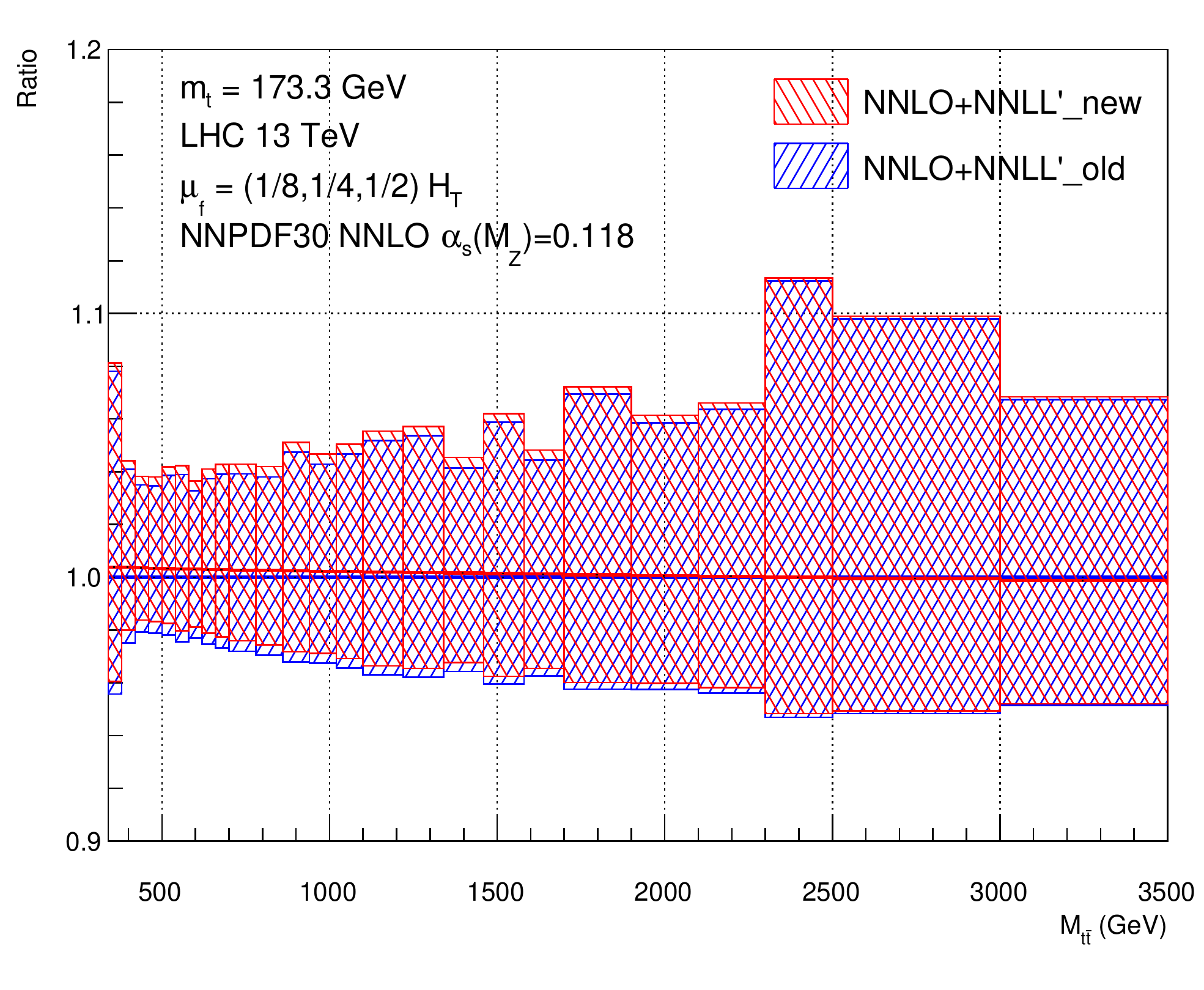}
\includegraphics[width=0.495\textwidth]{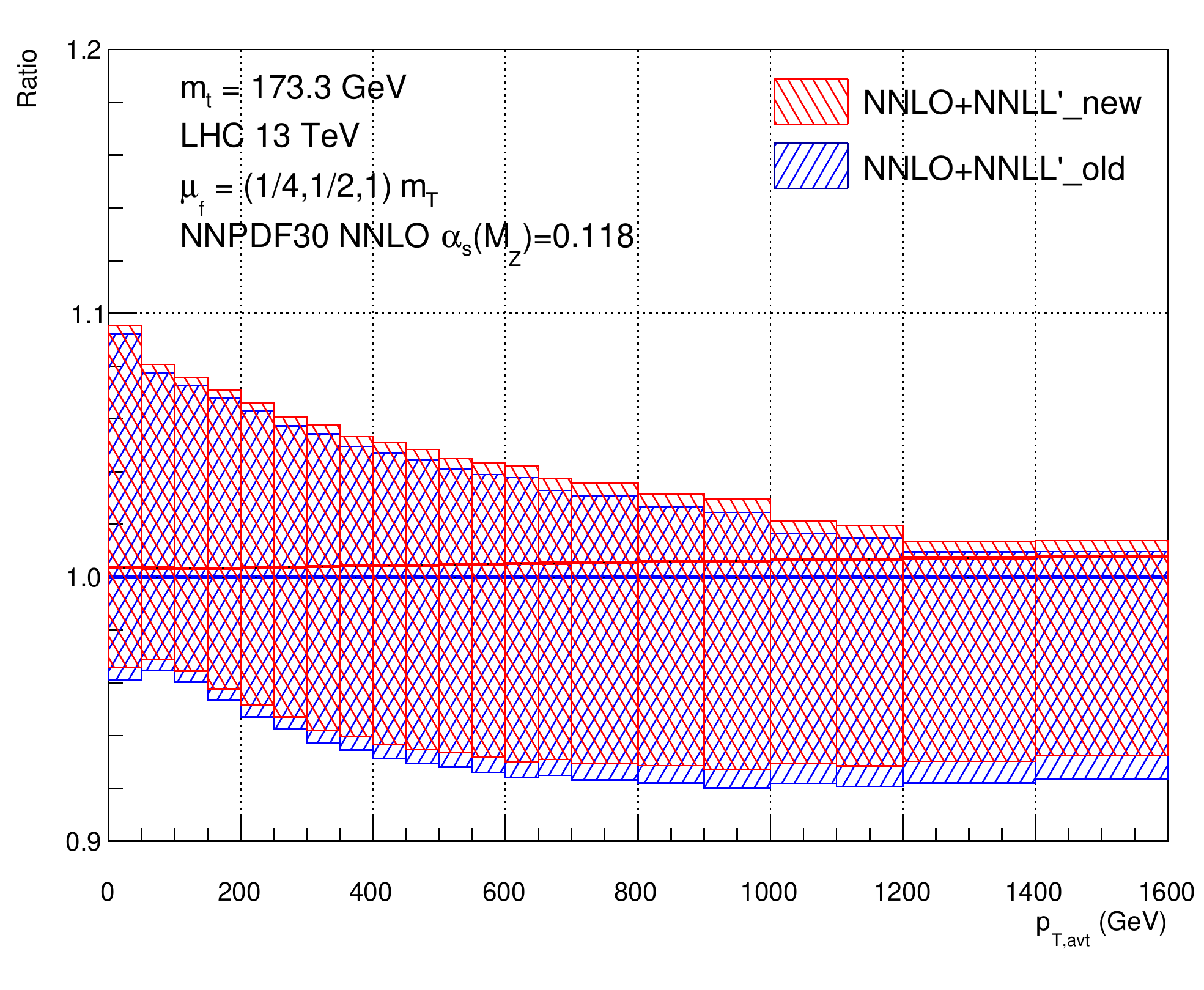}
\caption{\label{fig:cDsD3p}Comparison of the numerical results computed with the old (used in \cite{Pecjak:2016nee, Czakon:2018nun}) and the new (used in this work) coefficient functions $\widetilde{\bm{s}}_{ij}$, $C_D$ and $\widetilde{s}_D$. The default scale choices are given in \cite{Czakon:2018nun}. The results here are obtained using the NNPDF3.0 PDF set with ${\alpha_s(M_Z)=0.118}$ \cite{Ball:2014uwa}.}
\end{figure}

We now turn to assess the numerical impact of these modifications. Since the rapidity distributions have never been computed with the ``old'' coefficients, we only need to compare the $M_{t\bar{t}}$ and $p_{T,\text{avt}}$ distributions. In table~\ref{tab:cDsD}, we list the central values for the integrated cross section in two sample bins for each of these two distributions. The two bins are representative for the un-boosted and the boosted region, respectively. In figure~\ref{fig:cDsD3p}, we show the two distributions with uncertainty bands reflecting the scale variations. From both the table and the plots, we find that the numerical differences between the old and the new results are at the sub-percent level. We hence conclude that the modifications of the coefficient functions lead to no visible changes to the results presented in \cite{Pecjak:2016nee, Czakon:2018nun}.

\end{document}